\documentclass[aps,prd,twocolumn,numerical,superscriptaddress,floatfix,nofootinbib]{revtex4-2}
\usepackage{float,ulem}
\usepackage{graphicx,epsfig,epstopdf,amsmath,amssymb}
\usepackage{xcolor}
\usepackage[colorlinks=true,linkcolor=blue,citecolor=blue,urlcolor=blue]{hyperref}
\usepackage{caption}
\usepackage{subcaption}

\newcommand{\be}{\begin{equation}}
\newcommand{\ee}{\end{equation}}
\newcommand{\ba}{\begin{eqnarray}}
\newcommand{\ea}{\end{eqnarray}}
\newcommand{\nn}{\nonumber\\}

\begin{document}

\title{Acausality in truncated `MIS' - type theory}


\author{Sukanya Mitra}
\email{sukanya.mitra@niser.ac.in}
\affiliation{School of Physical Sciences, National Institute of Science Education and Research, An OCC of Homi Bhabha National Institute, Jatni-752050, India.}

\begin{abstract}
The causal-stable Muller-Israel-Stewart (MIS) theory is known to have a finite number of out of equilibrium derivative order corrections but requires treating the viscosity tensor as a separate degree of freedom with its own equations of motion, apart from the fundamental fluid degrees of freedom like velocity and temperature. In this work, I will show that it is possible to rewrite the MIS theory only in terms of velocity and temperature, but the resultant constitutive relation for dissipation must include all orders of gradient corrections. In this work I will argue that an all-order resummation of gradient contributions is equivalent to introducing new `non-fluid' degrees of freedom in the MIS theory. It will also be shown, using the relativistic quantum causality condition, that any finitely truncated order of derivative correction, however high it is, leads to a theory that is acausal, unless the corrections are infinitely summed up to all orders.
\end{abstract}
\maketitle

\section{Introduction}
Fluid dynamics is an effective theory that describes the dynamics of a near equilibrium system by the evolution of the conserved fields at long wavelength limit \cite{Landau,Kadanoff:1963axw}. The expectation values of these fields at equilibrium define the hydrodynamic state variables which serve as the fundamental fluid degrees of freedom. The out of equilibrium dynamics
is described in terms of the order by order gradient corrections of these fields. For a relativistic system however, the fluid theory needs to qualify some physical criteria such as causal wave propagation and stability against small perturbations.
The relativistic version of the long established  first order theories \cite{Landau,Eckart:1940te} turns  out to suffer from the pathologies regarding super luminal signal propagation and thermodynamic instability. As a rescue, the higher order Muller-Israel-Stewart (MIS) theory \cite{Muller,Israel:1976tn,Israel:1976efz,Israel:1979wp} has been suggested
to apply as the standard theory of relativistic dissipative fluid dynamics \cite{Hiscock:1983zz,Hiscock:1985zz,Olson:1990rzl}.

The MIS theory and the recent derivations of other analogous higher order theories \cite{Baier:2007ix,Denicol:2012cn,Muronga:2001zk,Jaiswal:2013vta} can serve as causal as well as stable hydrodynamic formalism with the fundamental fluid variables like velocity and temperature being locally fixed to their equilibrium values by the Landau gauge condition ($T^{\mu}_{\nu}u^{\nu}=-\varepsilon u^{\mu}$, $T^{\mu\nu}$ is the energy-momentum tensor, $\varepsilon$ and $u^{\mu}$ are the energy density and fluid velocity at their local equilibrium values). The price is paid by the limitation that these fundamental hydro fields are no longer sufficient to describe the fluid dynamics and additional degrees of freedom have to be introduced.
Hence, the dissipative fields in the MIS theory are promoted as the new degrees of freedom and are attributed their individual equations of motion. Although this version of relativistic hydrodynamic theories has been quite popular and phenomenologically successful
especially in the context of high energy heavy ion data analysis \cite{Huovinen:2006jp}, the physical meaning of these new degrees of freedom still remains somewhat questionable. These quantities do not relate to any conserved fields. In other words, they do not have any equilibrium counterparts.
Given the scenario, one might wonder if it is possible to have an equivalent ``fluid dynamical'' description of MIS theory where the requirement of these non-fluid new degrees of freedom can be eliminated. By ``fluid dynamical'' here I mean that, (i) the constitutive relation of the stress tensor can be entirely written in the terms of temperature, velocity and their derivations and the only equations of motion are the stress tensor conservation, and (ii) the fluid variables such as velocity and temperature are fixed by the Landau gauge such that there is no ambiguity in their definition.

Motivated by this idea, in this work, I attempt to rewrite the MIS theory that obey causality and stability while remaining in the Landau frame (such that the field variables are well defined) but without requiring any additional degrees of freedom. For simplicity, here a conformal system is considered without any conserved charges. We will see that it is indeed possible to generate the identical results of MIS theory only using the velocity and temperature as fluid variables, if the constitutive relation for the dissipation (here the shear tensor) runs up to infinite orders of gradient correction.
In this context, we remember that an all-order resummation of gradient contributions is known to be equivalent to introducing non-hydrodynamic modes which have been implemented in many ways. In \cite{Lublinsky:2009kv,Bu:2014sia}, the transport coefficients are resummed by making them frequency dependent in order to incorporate all order gradient corrections. In \cite{Heller:2013fn,Romatschke:2017vte}, these large orders are dealt with using Borel resummation techniques with Pade approximation which are also leading works of the far from equilibrium attractor theory. Muller-Israel-Stewart (MIS) is one such resummation scheme where the resummation is shown to result in not only non-hydro modes but new `non-fluid' degrees of freedom. I will then explicitly prove, using the causality condition ${\text{Im}}(\omega(k))\leq|{\text{Im}}(k)|$ of relativistic quantum theory \cite{Itzykson}, that any finite order truncation of the derivative correction series immediately leads to acausality unless the dissipations are promoted as individual degrees of freedom or summed infinitely.

Throughout the manuscript, I use natural unit~($\hbar = c = k_{B} = 1 $) and flat space-time with mostly positive metric signature $\eta^{\mu\nu} = \text{diag}\left(-1,1,1,1\right)$.
\section{MIS theory - An all order gradient correction theory}
To set the chain of arguments, I start with the well known form of the the MIS equations of motion keeping up to the linear terms \cite{Romatschke:2009im},
\begin{align}
&\partial_{\mu}T^{\mu\nu}=0~,~~T^{\mu\nu}=\varepsilon\left(u^{\mu}u^{\nu}+\frac{1}{3}\Delta^{\mu\nu}\right)+\pi^{\mu\nu}~,
\label{MIS1}\\
&\pi^{\mu\nu}+\tau_{\pi}D\pi^{\mu\nu}=-2\eta\sigma^{\mu\nu} ~.
\label{MIS2}
\end{align}
The conformal and uncharged energy-momentum tensor $T^{\mu\nu}$ consists of the equilibrium fields such as energy density $\varepsilon$ and hydrodynamic velocity $u^{\mu}$ (along with projection operator $\Delta^{\mu\nu}=\eta^{\mu\nu}+u^{\mu}u^{\nu}$) and the viscous correction $\pi^{\mu\nu}$ that depends upon shear viscosity $\eta$ and relaxation time of shear viscous flow $\tau_{\pi}$.
The notation $D (=u^{\mu}\partial_{\mu}  \text{~in local rest frame})$ indicates the temporal derivative correction of the hydrodynamic fields and $\sigma^{\mu\nu}=\Delta^{\mu\nu}_{\alpha\beta}\partial^{\alpha}u^{\beta}$ denotes the traceless, symmetric velocity gradient with $\Delta^{\mu\nu\alpha\beta}=\frac{1}{2}\Delta^{\mu\alpha}\Delta^{\nu\beta}+\frac{1}{2}\Delta^{\mu\beta}\Delta^{\nu\alpha}
-\frac{1}{3}\Delta^{\mu\nu}\Delta^{\alpha\beta}$.
Here I attempt to derive the combined results of Eq.\eqref{MIS1} and \eqref{MIS2} without treating $\pi^{\mu\nu}$ as an independent degree of freedom. Instead of attributing an individual differential equation to $\pi^{\mu\nu}$ like Eq.\eqref{MIS2}, I express it as a sum of order by order gradient corrections in Eq.\eqref{MIS1} itself as,
\begin{align}
&\pi^{\mu\nu}=\sum_{n}\pi_n^{\mu\nu}~,\nn
&\pi^{\mu\nu}_1=-2\eta\sigma^{\mu\nu}~,~~~
\pi_n^{\mu\nu}=-\tau_{\pi}D\pi^{\mu\nu}_{n-1}~,n \geq 2~.
\label{pisum}
\end{align}
This summation of order by order gradient correction in \eqref{pisum} leads to the shear-stress tensor as the following,
\begin{align}
 \pi^{\mu\nu}&=-2\eta\left\{\sum_{n=0}^{N}\left(-\tau_{\pi}D\right)^n\right\}\sigma^{\mu\nu}~,
 \label{MIS3}
\end{align}
where upto $N^{\text{th}}$ order of temporal derivative corrections have been taken.
 Next, I linearize the conservation equations (\eqref{MIS1} and \eqref{MIS2}) for small perturbations of fluid variables around their hydrostatic equilibrium as $\psi(t,x)=\psi_0+\delta\psi(t,x)$. Here, the subscript $0$ indicates global equilibrium and the fluctuations $\delta\psi(t,x)$ are expressed in the plane wave solutions via a Fourier transformation $\delta\psi(t,x)\rightarrow e^{i(kx-\omega t)} \delta\psi(\omega,k)$, with wave 4-vector $k^{\mu}=(\omega,k,0,0)$. Following the linearization, the shear and sound channel dispersion relations from Eq.\eqref{MIS3} respectively become,
\begin{align}
&(i\omega)+\tilde{\eta}(ik)^2\left[\sum_{n=0}^{N}\left(\tau_{\pi}i\omega\right)^n\right]=0~,
\label{disp-shear}\\
&(i\omega)^2+\frac{4}{3}\tilde{\eta}(i\omega)(ik)^2\left[\sum_{n=0}^{N}(\tau_{\pi}i\omega)^n\right]-\frac{1}{3}(ik)^2=0~.
\label{disp-sound}
\end{align}
Here $\tilde{\eta}=\eta/(\varepsilon_0+P_0)=\eta/(\frac{4}{3}\varepsilon_0)$ for a conformal system.

It can now be readily checked that if in Eq.\eqref{MIS3}, \eqref{disp-shear} and \eqref{disp-sound}, the sum over the gradient series is taken upto all orders ($N\rightarrow\infty$), the infinite sum results into a closed form, such as $\sum_{n=0}^{\infty}x^n=\frac{1}{1-x}$ with $x=\tau_{\pi}i\omega$. The sum exists within the radius of convergence $|\tau_{\pi}i\omega|<1$, whose circumference is the location of the first non-hydro mode of MIS theory ($\omega=-\frac{i}{\tau_{\pi}}$), beyond which hydro gradient expansion should not be trusted anyway. Applying this technique, Eq.\eqref{MIS3} turns out to be,
\begin{align}
 \pi^{\mu\nu}=-2\eta \left(1+\tau_{\pi}D\right)^{-1}\sigma^{\mu\nu}~,
 \label{MIS4}
 \end{align}
  where this infinite sum appears in the denominator of the expression of $\pi^{\mu\nu}$ in the form of the relaxation operator $\left(1+\tau_{\pi}D\right)$. We can see Eq.\eqref{MIS4} readily takes us to Eq.\eqref{MIS2}. Hence, we see that the all order infinite sum in $\pi^{\mu\nu}$
in Eq.\eqref{pisum} is producing the identical results of solving the $\pi^{\mu\nu}$ from Eq.\eqref{MIS2} by considering it an independent degree of freedom, without actually doing so. Consequently, from Eq.\eqref{disp-shear} and \eqref{disp-sound} with $N\rightarrow\infty$, we find the well known dispersion polynomials of MIS theory \cite{Pu:2009fj}, which at local rest frame for shear and sound channels are given respectively by:
\begin{align}
 &\tau_{\pi}\omega^2+i\omega-\tilde{\eta}k^2=0~,
 \label{MIS-Shear}\\
 &\tau_{\pi}\omega^3+i\omega^2-\left(\frac{4}{3}\tilde{\eta}+\frac{1}{3}\tau_{\pi}\right)\omega k^2-\frac{1}{3}ik^2=0~.
\end{align}
So if we choose to integrate out $\pi^{\mu\nu}$ from Eq.\eqref{MIS2} by solving it first in terms of the fluid variables using the perturbative technique of derivative expansion, the resultant fluid equations turn out to have an infinite number of derivatives (Eq.\eqref{pisum} with $N\rightarrow\infty$).
This infinite sum over the gradient corrections (Eq.\eqref{pisum}) being `integrated in' (summed in a closed form to generate temporal derivatives in the denominator such that $\pi^{\mu\nu}$ gets its own differential equation \eqref{MIS4}) as the new degrees of freedom in MIS theory, is the key result of the current work.

Based on these results, in this work I argue that, the theory well known as the MIS theory
(Eq.\eqref{MIS1} and \eqref{MIS2} combined) that is known to be free from the acausal signal propagation and thermodynamic instabilities, is not a second order or any finite higher order truncated theory. Rather an infinite sum over the derivative order corrections is required to produce Eq. \eqref{MIS1} and \eqref{MIS2}. This infinite sum is actually equivalent to attributing a new degree of freedom other than velocity and temperature. Describing the dissipative dynamics of a relativistic system in terms of gradient corrections
renders a pathology free theory only and only if the corrections are summed up to infinite orders. Any truncation at finite order directly leads to the violation of causality and can not serve as an acceptable hydrodynamic theory.

  In the following sections, I will establish that any finitely truncated order summation of $\pi^{\mu\nu}$ in Eq.\eqref{pisum}, however high it is, can not produce a pathology free stable-causal theory. In order to resolve the issues of causality and stability, we need to consider all orders of gradient corrections in Eq.\eqref{MIS3} ($N\rightarrow\infty$), such that the infinitely summed derivatives generate the relaxation operator $(1+\tau_{\pi}D)$ in the denominator of $\pi^{\mu\nu}$. To prove this, I take recourse of the relativistic quantum theory causality condition,
\be
\text{Im}(\omega(k))\le |\text{Im}(k)|~,
\label{Heller}
\ee
which indicates the stability invariance of a theory as well \cite{Heller:2022ejw,Gavassino:2023myj,Hoult:2023clg}. In the following, I will show case by case that a truncated theory always violates Eq.\eqref{Heller}, while an all order theory resulting from the infinite sum restores the causality and stability by always satisfying \eqref{Heller} for $\tau_{\pi}>\tilde{\eta}$.
\section{Testing truncated theory}
In order to check condition \eqref{Heller}, we need to extract the solution $\omega(k)$ from the respective dispersion polynomial. Here as a test case, I investigate the shear channel frequency solution under the inequality \eqref{Heller}. As mentioned in the previous section, for a truncated theory where $\pi^{\mu\nu}$ is taken upto a certain finite order $N$, the shear channel dispersion polynomial \eqref{disp-shear} becomes :
\begin{align}
(i\omega)+\tilde{\eta}(ik)^2\left[\sum_{n=0}^{N}\left(\tau_{\pi}i\omega\right)^n\right]=0~.
\label{disp-shear-trunc}
\end{align}
I will now show that for any finite $N$, the solution of Eq.\eqref{disp-shear-trunc} is not able to satisfy the condition given in \eqref{Heller}
for all possible values of $k$ \footnote{This part of the calculation has been done with the collaboration of Sayantani Bhattacharyya.}.

{\subsection{Truncation at $N=0$}}
For $N=0$, we have the usual Navier-Stokes (N-S) shear channel which has an exact solution, $\omega=-i\tilde{\eta}k^2$. For any real $k$, the condition \eqref{Heller} requires
$\tilde{\eta}$ to be positive. But if $k$ is a purely imaginary large number $k=\pm ip$, with $p(>>1)$ real and positive, we have $\text{Im}(\omega)=\tilde{\eta}p^2$, which with $\tilde{\eta}>0$ clearly violates $\text{Im}(\omega)\le |\text{Im}(k)|$ since $\text{Im}(\omega)\sim p^2$ and $|\text{Im}(k)|\sim p$.

{\subsection{Truncation at $N=1$}}
For $N=1$, we have the exact solution,
\be
\omega=-i\tilde{\eta}k^2/\left[1-\tilde{\eta}\tau_{\pi}k^2\right]~.
\ee
For a value of $k$ given by, $\tilde{\eta}\tau_{\pi}k^2=(1\pm\epsilon)$ with $\epsilon$ to be real, positive and $\epsilon << 1$, $\text{Im}(\omega_{\pm})=\pm\frac{1}{\tau_{\pi}}\frac{1}{\epsilon}$ which is a large number. Now if $\tau_{\pi}$ is positive ($\eta$ is also a positive number \cite{Weinberg:1971mx} considering the constraints of the second law of thermodynamics), then $k$ is a real number with $\text{Im}(k)=0$ and $\text{Im}(\omega_+)=1/(\tau_{\pi}\epsilon)$ is a large positive number. So $\text{Im}(\omega)\le |\text{Im}(k)|$ is violated. If $\tau_{\pi}$ is negative, $\text{Im}(\omega_-)=-1/(\tau_{\pi}\epsilon)$ is again a large positive number with $|\text{Im}(k)|=1/\sqrt{\tilde{\eta}|\tau_{\pi}|}$ which is a finite quantity.
So $\text{Im}(\omega)\le |\text{Im}(k)|$
is again violated.

{\subsection{Finite truncation at $N\geq 2$}}
For $N\geq 2$ but still with finite truncation, Eq.\eqref{disp-shear-trunc} is only possible to solve at limiting values of $k$.
At $k\rightarrow 0$, Eq.\eqref{disp-shear-trunc} gives at least one solution,
\begin{align}
(i\omega)=\left[\frac{1}{\tilde{\eta}\tau_{\pi}^N k^2}\right]^{\frac{1}{N-1}}.
\label{disp2}
\end{align}
Here the relaxation time and the wave vector is expressed as,
\be
\tau_{\pi}=|\tau_{\pi}|e^{i\sigma\pi}~,~~~~k=|k|e^{im\pi}~,
\ee
with $\sigma=0$ and $\sigma=1$ correspond to positive and negative values of relaxation time respectively, and $m$ is any real number. The imaginary part of frequency and wave number respectively becomes,
\begin{align}
&\text{Im}(\omega)=-\left[\frac{1}{\tilde{\eta}|\tau_{\pi}|^N |k|^2}\right]^{\frac{1}{N-1}}\text{cos}\left\{\frac{(2m+\sigma N)}{N-1}\pi\right\}~,\nn
&|\text{Im}(k)|=|k||\text{sin}(m\pi)|~.
\end{align}
We can always choose a domain such as $\frac{1}{2}<(2m+\sigma N)/(N-1)<\frac{3}{2}$, where the cosine function of $\text{Im}(\omega)$ is negative such that $\text{Im}(\omega)$ is a large positive number if $|k|$ is small. But $|\text{Im}(k)|$ is finite at small $|k|$. So clearly condition \eqref{Heller} is violated.

So we see that for any truncated value of $N$ starting from $0$, there always exists at least one mode such that the imaginary part of $\omega$ becomes greater than the modulus of imaginary part of $k$ in the complex $k$-plane and hence has issues with causality and stability. Thus we can safely conclude that any truncated order of gradient correction in Eq.\eqref{pisum}, can not produce a relativistic, dissipative theory that is consistent with the causality-stability assessment.
\section{Testing all order theory}
The all order infinitely summed theory has the shear channel dispersion polynomial as the following,
\be
\tau_{\pi}(i\omega)^2-(i\omega)-\tilde{\eta}(ik)^2=0~.
\label{disp-shear-all}
\ee
The solution of Eq.\eqref{disp-shear-all} is,
\begin{align}
 \omega_{\pm}=\frac{1}{2\tau_{\pi}}\left[-i\pm\left\{-1+4\tilde{\eta}\tau_{\pi}k^2\right\}^{\frac{1}{2}}\right]~.
 \label{MIS-sh-sln}
\end{align}
Now decomposing $\omega$ as $\omega=\omega_R+i\omega_I$ and expressing $k$ as $k=|k|e^{im\pi}$ as before, it is possible to derive,
\begin{align}
 &2\left(1+2\tau_{\pi}\omega_I\right)^2=\left\{1-4
 \tilde{\eta}\tau_{\pi}|k|^2\text{cos}(2m\pi)\right\}\nn
&~~~~~\pm\left[1+16\tilde{\eta}^2\tau_{\pi}^2|k|^4-8\tilde{\eta}\tau_{\pi}|k|^2\text{cos}(2m\pi)\right]^{\frac{1}{2}}~.
\label{Heller1}
\end{align}
In the following, it will be proved that for any value of $|k|$ and $m$, it is not possible to violate condition \eqref{Heller}. If possible, violation of \eqref{Heller} demands,
\be
\omega_I>|k||\text{sin}(m\pi)|~.
\label{Heller2}
\ee
Eq.\eqref{Heller1} and \eqref{Heller2} simultaneously lead us to the inequality,
\begin{align}
\pm&\left[\left\{1-4\tilde{\eta}\tau_{\pi}|k|^2\right\}^2+16\tilde{\eta}\tau_{\pi}|k|^2\text{sin}^2(m\pi)\right]^{\frac{1}{2}}>\nn
&1+4\tilde{\eta}\tau_{\pi}|k|^2+8\tau_{\pi}|k||\text{sin}(m\pi)|\nn
&+8\tau_{\pi}(\tau_{\pi}-\tilde{\eta})|k|^2\text{sin}^2(m\pi).
\label{Heller3}
\end{align}
We can see that, if $\tau_{\pi}>\tilde{\eta}$, the maximum value of the left hand side and the minimum value of the right hand side of \eqref{Heller3} both are $(1+4\tilde{\eta}\tau_{\pi}|k|^2)$. So the left hand side of \eqref{Heller3} can never exceed the right hand side and the violation of \eqref{Heller} is never possible once the condition $\tau_{\pi}>\tilde{\eta}$ is satisfied. Interestingly enough, this is the asymptotic causality condition of MIS shear channel estimated from the large wave number propagating mode, that can be obtained from \eqref{Heller}
itself.

So we conclude that, once the condition $\tau_{\pi}>\tilde{\eta}$ is obeyed by the transport coefficients, the causality condition $\text{Im}(\omega)\leq|\text{Im}(k)|$ is satisfied for all possible $k$ values in an all order theory. Hence, in contrast to the truncated ones, an all order theory
is the only admissible candidate for a reliable hydrodynamic theory for relativistic, dissipative fluids.

\section{Conclusion}
In this work it has been shown that though in the conventional MIS theory the equations of motion for temperature, velocity and viscous tensor have finite number of derivatives, integrating out $\pi^{\mu\nu}$ by solving it in terms of the fluid variables perturbatively, results into infinite number of derivatives. For such a theory, the
causal signal propagation is not compromised as long as the gradient corrections are summed up to infinite orders. The advantage of such an approach is that, no new degrees of freedom that could not be linked to conserved quantities are needed to be introduced, but still the temperature and velocity can be unambiguously fixed by the Landau gauge.
The recently proposed first-order stable and causal BDNK theory \cite{Bemfica:2017wps,Bemfica:2019knx,Bemfica:2020zjp,Kovtun:2019hdm,Hoult:2020eho,Hoult:2021gnb} does not require any additional degrees of freedom,
but they are well behaved only away from the Landau frame. Hence, the
primary field variables like velocity and temperature have ambiguities in their first principle definition since they do not coincide with those in either Landau’s or Eckart’s frames apart from global equilibrium and consequently lack a definition in terms of the microscopic field
theory operator $T^{\mu\nu}$.
Of course the theory derived here with infinitely many derivatives has practical limitations for simulation purpose since it is not possible to solve them even numerically for arbitrary initial conditions. But the point of this work is to indicate that if we want to construct a relativistic dissipative hydrodynamic theory purely in terms of fundamental fluid fields like temperature and velocity that have first principle microscopic definition, causality is only maintained if all orders of derivative corrections are taken into consideration.
If this all order sum needs to be averted,
it has to be folded in one way or another : either in terms of new degrees of freedom (MIS) or in terms of field redefinition leading to no first principle definition of velocity or temperature (BDNK).
In our recent article \cite{Bhattacharyya:2023srn} this tension has been further investigated. It has be shown that in BDNK theory if we want to define the velocity and temperature locally in terms of the stress tensor operator like we do in Landau frame, then the constitutive relation will include all orders of derivative corrections as well.

\section{Acknowledgements}
I duly acknowledge Sayantani Bhattacharyya for conceptual discussions and valuable inputs. I also acknowledge Shuvayu Roy for useful discussions.
For the ﬁnancial support S.M. acknowledges the Department of Atomic Energy, India.

\end{document}